\documentclass[aps,prb,twocolumn,superscriptaddress]{revtex4-2}
\usepackage{bm}
\usepackage{graphicx}
\usepackage{graphics}
\usepackage{amsmath}
\usepackage{amsfonts}
\usepackage{amssymb}
\usepackage{makeidx}
\usepackage[colorlinks=true,citecolor=blue,linkcolor=blue,linktocpage=true,pagebackref=false]{hyperref}
\hypersetup{colorlinks=true,citecolor=blue,linkcolor=blue,filecolor=blue,urlcolor=blue}
\usepackage{color,soul} 
\usepackage{float}
\usepackage{subfigure}
\usepackage{xcolor}

\usepackage[normalem]{ulem} % permite usar fonte tachada

\usepackage{IEEEtrantools}

\graphicspath{{figures/}} 
\begin{document}
\newcommand{\be}   {\begin{equation}}
\newcommand{\ee}   {\end{equation}}
\newcommand{\ba}   {\begin{eqnarray}}
\newcommand{\ea}   {\end{eqnarray}}
\newcommand{\ve}  {\varepsilon}
\newcommand{\Dis} {\mbox{\scriptsize dis}}
\newcommand{\state} {\mbox{\scriptsize state}}
\newcommand{\band} {\mbox{\scriptsize band}}
%%%%%%%%%%%%%%%%%%%%%%%%%%%%%%%%%%%%%%%%%%%%%%%%%%%%%%
\title{
Stoner ferromagnetism in low-angle twisted bilayer graphene at three-quarters filling}
\author{Kevin J. U. Vidarte}
\affiliation{Instituto de F\'{\i}sica, Universidade Federal 
Rio de Janeiro, 21941-972 Rio de Janeiro, RJ, Brazil}

\author{Felipe Pérez Riffo}
\affiliation{Departamento de Física, Universidad Técnica Federico Santa María, Casilla 110-V, Valparaíso, Chile}

\author{Eric Suárez Morell}
\affiliation{Departamento de Física, Universidad Técnica Federico Santa María, Casilla 110-V, Valparaíso, Chile}

\author{Caio Lewenkopf}
\affiliation{Instituto de F\'{\i}sica, Universidade Federal
Fluminense, 24210-346 Niter\'oi, RJ, Brazil}

\date{\today}
%----------------------------------------------------------------
\begin{abstract}
We present a theoretical investigation of the magnetic properties exhibited by twisted bilayer graphene (TBG) systems with small twist angles, where the appearance of flat minibands strongly enhances electron-electron interaction effects. 
We show that, at three-quarters filling of the conduction miniband, the Stoner mechanism induces a ferromagnetic polarization in the AA-stacking regions, which aligns with recent experimental observations.
Our approach models the electronic properties by a tight-binding Hamiltonian combined with a Hubbard mean-field interaction term. 
We employ a real-space recursion technique to self-consistently calculate the system's local density of states and use our method to investigate the magnetic properties of small-angle TBG at three-quarters filling.
% The latter allows us to compute the interaction term and solve the problem self-consistently. 
The recursion method's $O({\cal N})$ efficiency makes it possible to address extremely large superlattices through a full real-space approach. 
We validate our procedure by comparing it with mean-field momentum-space calculations from the literature, which identify a magnetic phase in charge-neutral TBGs. 
\end{abstract}
%------------------------------------------------------------------------
\maketitle

%%%%%%%%%%%%%%%%%%%%%%%%%%%%%%%%%%%%%%
\section{Introduction}
\label{sec:introduction}
%%%%%%%%%%%%%%%%%%%%%%%%%%%%%%%%%%%%%%

Twisting two layers of graphene by a specific angle has led to the discovery of a wealth of unexpected phenomena \cite{MacDonald2019, Andrei2020}. 
Particularly intriguing is the experimental discovery of insulating phases and superconducting states \cite{Cao2018a, Cao2018} in small-angle twisted bilayer graphene (TBG) systems, which has opened new paths for the investigation of graphene systems \cite{Yankowitz2019, Kerelsky2019, Lu2019, Serlin2020, Saito2021, Utama2021,  Lisi2021,  Tseng2022, Liu2022, Lin2022, Jaoui2022}.
This line of investigation, known as {\it twistronics} \cite{Carr2017}, has rapidly attracted a large and active community of researchers in condensed matter physics and materials science \cite{Andrei2020, Wang2019}.

The origin of these remarkable phenomena is attributed to the appearance of flat electronic bands in low-angle TGBs near the charge neutrality point (CNT) \cite{Morell2010, Laissardiere2010}, as explained by the continuum model \cite{Lopes-dos-Santos2007, Bistritzer2011, Lopes-dos-Santos2012, Tarnopolsky2019}.
The corresponding electronic states are localized, leading to enhanced electron-electron interaction effects.
% which originate the rich properties observed in low-angle TBGs. 
From a theoretical point of view, the challenge is to treat a system with a unit cell that has thousands of atoms with (strongly) interacting electrons.

The focus of our paper is to present a mean-field approach that deals with this issue, being capable of addressing systems with very large Hilbert spaces. 
As a showcase, we address the magnetic properties of low-angle TBG systems at finite doping.
The motivation is a recent experiment \cite{Sharpe2019} that observed the emergence of a ferromagnetic phase in a TBG with a twist angle $\theta \approx 1.16^\circ$ at 3/4 filling of the conduction miniband.
The experiment also observed a chiral edge state.
These findings have been nicely interpreted by the presence of flat bands having non-zero Chern numbers \cite{Zhang2019, Bultinck2020}, originated from the rotational alignment of the TBG to the hBN substrate that gives rise to band topological effects, and originated from the flat bands having non-zero Chern numbers. 
Our study does not challenge this theory. 
We rather add another possible mechanism to explain the ferromagnetic phase. 

The onset of magnetism in graphene-based systems has been intensively studied, both experimentally and theoretically \cite{Yazyev2010}. 
The latter span a wide variety of settings, such as zero-dimensional graphene nanofragments \cite{FernandezRossier2007,Wang2009,Feldner2010}, one-dimensional graphene nanoribbons \cite{Fujita1996, Son2006PRL, FernandezRossier2008, Pisani2007, Tao2011,  Carvalho2014}, defect-induced magnetism in bulk graphene \cite{Yazyev2007,Miranda2016}.
The description of magnetic moments induced by edge terminations and vacancies has been addressed by both {\it ab initio} calculations and the tight-binding approximation with an on-site Hubbard electron-electron interaction term. 
The magnetic properties of these systems are nicely described by a mean-field theory, with the exception of vacancy-induced Kondo correlations \cite{Chen2011, Miranda2014, Jiang2018}.  

In this paper, we employ a tight-binding Hamiltonian with a mean-field Hubbard on-site interaction term to compute the low-energy electronic and magnetic properties of low-angle TGBs.
In particular, we study the Stoner magnetization at  $3/4$-filling of the conduction miniband of TBGs with $\theta \approx 1.16^\circ$.  
We use the Haydock-Heine-Kelly (HHK) recursive technique \cite{Vidarte2022,Haydock1972,Haydock1975,Haydock1980} to calculate the spin-resolved LDOS 
% with spin $\sigma = \{ \uparrow ,\downarrow \}$ 
as a function of the twist angle $\theta$. 
Being an order $\mathcal{N}$ method, the HHK method makes possible to compute the Green’s functions of TBG with very large primitive unit cells. 
The latter combined with a self-consistent mean-field calculation allows us to investigate the electron localization properties and the magnetization of TBGs at arbitrary filling factors of the flat minibands.
% Motivated by a recent experimental paper \cite{Sharpe2019}, our focus is the study of emergent ferromagnetism close to $3/4$-filling at the conduction miniband of a TBG with $\theta \approx 1.16^\circ$ \cite{Sharpe2019}. 

The paper is organized as follows. 
In Sec.~\ref{sec:method} we review the geometric properties of commensurate TBG moir\'e structures, present the Hamiltonian model, and the numerical method developed for this study. 
In Sec.~\ref{sec:results} we discuss the LDOS of low-angle TBG rigid and relaxed strucutures. 
First, we show our results for non-interacting electrons. 
Next, we calculate the magnetic properties of TBGs at the charge neutrality point (CNP). 
The excellent agreement between our results and those of Ref.~\cite{Vahedi2021}, discussed in App.~\ref{Appendix}, serves to validate our method. 
Finally, we investigate the magnetic properties of a TBG system with $\theta=1.16^\circ$ at 3/4 filling and find the emergence of a ferromagnetic phase, in line with recent experiments \cite{Sharpe2019}. 
We present our conclusions in Sec.~\ref{sec:conclusion}.

%%%%%%%%%%%%%%%%%%%%%%%%%%%%%%%%%%%%%%
\section{Theory and methods}
\label{sec:method}
%%%%%%%%%%%%%%%%%%%%%%%%%%%%%%%%%%%%%%

\subsection{TBG geometric and symmetry properties}
\label{sec:geometry}

The stacking geometry of TBGs is characterized, as standard \cite{Lopes-dos-Santos2007, Rozhkov2016, katsnelson2020}, by the twist of one graphene layer with respect to the other around a given site starting from the AA-stacked bilayer, forming a moir\'e pattern.
We describe a TBG system by twisting the upper layer at an angle $\theta$ as described in Ref.~\cite{Lopes-dos-Santos2007,Bistritzer2011}.
The primitive lattice vectors of the bottom layer are $\textbf{a}^{b}_{1}=\sqrt{3} a_0 \hat{\bf e}_{x}$ and $\textbf{a}^{b}_{2}=\sqrt{3}a_0/2 \left( \hat{\bf e}_{x} + \sqrt{3}\hat{\bf e}_{y} \right)$ with the carbon-carbon bond length $a_{0}=1.42$ \AA, and those of top layer as $\textbf{a}^{t}_{i}=R(\theta)\textbf{a}^{b}_{i}$, where $R(\theta)$ is the rotation matrix.
In this study, we take the spacing between graphene layers as $d_{0} = 3.35$~\AA~for non-relaxed lattices \cite{Laissardiere2010}.   

The lattice structure of a TBG system is commensurate if the periods of the two graphene layers match, giving a finite unit cell. 
Hence, the periodicity condition requires the lattice translation vector $m\textbf{a}_{1}^{b} + n \textbf{a}_{2}^{b}$ of the bottom (unrotated) layer and $n\textbf{a}_{1}^{t} + m\textbf{a}_{2}^{t}$ in the top (rotated) layer, with $m$ and $n$ integers, to coincide. 
The twist angle $\theta$ for a commensurate structure is related to $(m,n)$ by \cite{Moon2013,Mele2010,Mele2012} 
\be
\theta (m,n)= 	\arccos{\left( \dfrac{1}{2}\dfrac{m^{2}+n^{2}+4mn}{m^{2}+n^{2}+mn} \right) } ,
\ee
with a lattice constant
\be
\label{Eq:lattice_constant}
L=a_{0}\sqrt{3(m^{2}+n^{2}+mn)}=\dfrac{\vert m-n\vert\sqrt{3}a_{0}}{2\sin{\theta/2}}.
\ee
The commensurate unit cell contains $N=4(m^{2}+n^{2}+mn)$ atoms.

Due to the symmetry of the honeycomb lattice, when the translation vector of the top layer is fixed to $\boldsymbol{\delta}^{t}=0$, a twist at an angle $\theta =120^{\circ}$ transforms the AA-stacked bilayer into itself. 
In turn, a twist by $\theta = 60^{\circ}$ transforms the bilayer from AA- to AB-stacking. 
TBGs with $\theta$ and $-\theta$ are mirror images that share equivalent band structures \cite{Moon2013}.

\begin{figure}[h!]
\includegraphics[width=0.85\columnwidth]{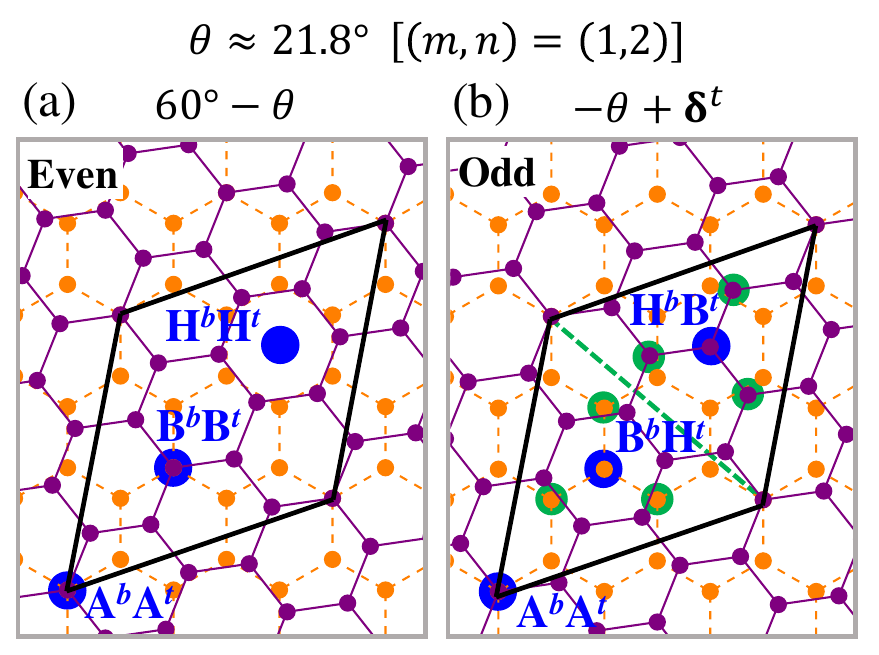}
\caption{Commensurate structure partners with $\theta \approx 21.8^{\circ}$ that corresponds $(1,2)$.
Purple (orange) circles indicate the carbon sites of the top (bottom) layer.
The black lines indicate the primitive unit cell.
The 3 blue disks correspond to the three-fold symmetry positions.
(a) The twist angle $60^{\circ} -\theta$ corresponds to an even symmetry under sublattice exchange at three-fold symmetric positions.
(b) The twist angle $-\theta$ followed by $\boldsymbol{\delta}^{t}$ translation represents an odd symmetry under sublattice exchange.
The green disks indicate the equivalent points. 
The green dashed line corresponds to the two-fold rotation axis.
}
\label{fig:commesurate}
\end{figure}

Each graphene layer is constituted by two sublattices, ${\rm A}^{b({\rm or}\;t)}$ and ${\rm B}^{b({\rm or}\;t)}$.
The commensurate structures occur in two different forms distinguished by their sublattice parities \cite{Mele2010,Mele2012}. 
Figure \ref{fig:commesurate}(a) shows an example of an even commensurate structure under sublattice exchange. 
It is characterized by having 3 three-fold symmetric positions that correspond to the stacking of the ${\rm A}^{b}{\rm A}^{t}$ and ${\rm B}^{b}{\rm B}^{t}$  sites and by hexagonal centers that overlap, ${\rm H}^{b}{\rm H}^{t}$. 
In contrast, Fig. \ref{fig:commesurate}(b) shows an odd commensurate structure.
Here, the top and bottom coinciding sites correspond to ${\rm A}^{b}{\rm A}^{t}$ at the origin, and the two remaining three-fold symmetric positions are occupied by ${\rm B}^{b}$(or ${\rm B}^{t}$)-sublattice sites of one layer aligned with the hexagon centers ${\rm H}^{t}$(or ${\rm H}^{b}$) of its neighbor layer. 
Therefore, the angles $60^{\circ} -\theta$ and $-\theta$ followed by a relative translation of the upper layer by $\boldsymbol{\delta}^{t} = ( \textbf{a}_{1}^{t} + \textbf{a}_{2}^{t} )/3$ form commensurate partners with unit cells of equal areas but opposite sublattice parities. 
 
%--------------------------------------- F I G U R E -------------------------------------------
%------------------------------------------------------------------------------------------	

A TBG with $\theta=30^{\circ}$ is a special case, since its crystal structure
is its own commensurate partner and corresponds to an elementary two-dimensional quasicrystalline lattice \cite{Yao2018,Ahn2018,Vidarte2024}.

Here, we work with odd commensurate structures. 
In Fig. \ref{fig:commesurate}(b), the carbon atom sites, indicated by the green disks, are equivalent due to the three-fold symmetric positions and two-fold rotation axis (green dashed line). 
These symmetry properties allow us to reduce the numerical computation by identifying the equivalent sites within the primitive unit cell.

Lattice relaxations have been shown to significantly influence the electronic states of low-angle TBGs \cite{Nam2017}.
Our calculations, which incorporate both in-plane and out-of-plane lattice relaxations, are implemented as follows: 
For a given twist-angle $\theta$ we determine the atomic relaxation positions of the lattice structures defined above using the Large-scale Atomic/Molecular Massively Parallel Simulator (LAMMPS) \cite{lammps_1995}. The energy minimization process involves both non-bonded and bonded interactions to accurately model the system.
Non-bonded interlayer interactions are handled using a registry-dependent interlayer potential specifically designed for graphene/hBN heterostructures \cite{Leven2014,Leven2016,Maaravi2017}, with a cutoff distance of 16 \AA, ensuring that all relevant interlayer forces are considered. The parameters for the registry-dependent potential are given in \cite{Ouyang2018}. 
Bonded intralayer interactions are modeled using the Tersoff potential \cite{Tersoff1988,Kinaci2012}, which is well-suited for describing the covalent bonding in carbon-based materials like graphene.  
The lattice structure relaxation calculations are performed with an energy tolerance criterion set to $10^{-10}$ eV per atom  \cite{Mostofi_2020}.

%--------------------------------------------------------------------
\subsection{Model Hamiltonian}
\label{sec:ModelHamiltonian}
%--------------------------------------------------------------------

We model the electronic properties of low-angle TBG systems using the effective Hamiltonian
%a tight-binding Hamiltonian with a Hubbard interaction term \cite{Yazyev2010,Auerbach1998}, namely
\be
\label{Eq:Hamiltonian}
    H = H_{\rm TB} + H_{\rm U},
\ee
where $H_{\rm TB}$ stands for the TBG tight-binding Hamiltonian and $H_{\rm U}$ for a Hubbard on-site electron-electron interaction term \cite{Yazyev2010,Auerbach1998}.

The tight-binding Hamiltonian reads
% that describes the low-energy electronic structure of TBG reads
\be
\label{eq:H_tb}
H_{\rm TB} = \sum_{i,j\sigma} \left(  t_{ij} c^\dagger_{i\sigma} c^{}_{j\sigma} + {\rm H. c} \right) ,
\ee
where the operators $c^{}_{i\sigma}$ and $c^\dagger_{i\sigma}$ annihilate and create an electron with spin $\sigma = \left\lbrace \uparrow , \downarrow \right\rbrace $ at site $i$, respectively.
$t_{ij}$ is the transfer integral between the Wannier electronic orbitals centered at the carbon sites $i$ and $j$. 
The transfer integral $t_{ij} \equiv t({\mathbf R}_{ij})$, where ${\mathbf R}_{ij} ={\bf R}_i - {\bf R}_j$, depends on the interatomic distance and on the relative orientation between $p_{z}$ orbitals. 
$t_{ij}$ is parameterized as \cite{Laissardiere2010,Moon2013,Uryu2004}
\be
\label{eq:Hopping_energy}
t({\mathbf R}) = V_{pp\pi} (R)  \left[ 1-\left( \frac{ {\mathbf R} \cdot  \mathbf{e}_z } {R} \right)^2 \right] +  V_{pp\sigma} (R) \left( \frac{ {\mathbf R} \cdot  \mathbf{e}_z } {R} \right)^2\!,
\ee
with
\begin{equation}
 V_{pp\pi} (R) =  V^0_{pp\pi} \exp \left( -\frac{R - a_0}{ r_0} \right) 
\end{equation}
\begin{equation}
 V_{pp\sigma} (R) =  V^0_{pp\sigma} \exp \left( -\frac{R - d_0}{ r_0} \right),  
\end{equation}
% where $R=\vert \textbf{R}_{i} - \textbf{R}_{j} \vert ^{1/2}$. 
Here, $V^{0}_{pp\pi}=-2.7$~eV, 
% is the transfer integral between nearest-neighbor atoms within a graphene layer, 
$V^{0}_{pp\sigma}=0.48$~eV,
% is the interlayer transfer integral between vertically located atoms
and $r_0 =0.319 a_0$ is the decay length of the transfer integral \cite{Uryu2004,Laissardiere2010}. 
The transfer integral for $R > 5 $~\AA~is exponentially small and can be safely neglected.
The intra- and interlayer hopping matrix elements $t_{ij}$ have been determined by fitting the dispersion relations of graphene monolayer and graphene AB-stacked bilayer obtained from {\it ab initio} calculations \cite{Laissardiere2010}.

The Hubbard term reads
\be
    H_{\rm U} = U \sum_{i} n^{}_{i\uparrow} n^{}_{i\downarrow} ,
\ee
where $n^{}_{i\sigma} = c^\dagger_{i\sigma} c^{}_{i\sigma}$  is the spin-resolved electron number operator at site $i$.
The parameter $U > 0$ gives the magnitude of the on-site Coulomb repulsion \cite{Auerbach1998,Yazyev2010}.
The magnitude of $U$ in graphene systems is has been extensively discussed \cite{Hancock2010, Schuler2013} with estimates that vary from $0.5 \cdots 2.0 V^{0}_{pp\pi}$. 
In this study we consider, unless otherwise stated, $U = V^{0}_{pp\pi}$. 

We solve the TBG Hamiltonian for a given filling factor in the mean-field approximation.
As standard, we write $n_{i\sigma} \equiv \langle n_{i\sigma}\rangle + \delta n_{i\sigma}$ and neglect the quadratic terms in $\delta n_{i\sigma}$ that are responsible for electronic correlations. 
The mean field Hubbard Hamiltonian reads
\be
    H_{\rm U}^{\rm MF} = U \sum_{i} \left[ n^{}_{i\uparrow} \langle n_{i\downarrow}\rangle + n^{}_{i\downarrow} \langle n_{i\uparrow}\rangle 
                - \langle n_{i\uparrow}\rangle \langle n_{i\downarrow}\rangle \right] .
\ee
The last term is a constant for a given local magnetic configuration and can be absorbed in the chemical potential.

%--------------------------------------------------------------------
\subsection{Numerical method}
\label{sec:Numericalmethod}

Let us now describe how the computation of the ground state charge density and magnetic properties are implemented. 
%of TBG systems modeled by the Hamiltonian $H=H_{\rm TB} + H_{\rm U}^{\rm MF}$.

Due to the large number of atoms in a low-angle TBG primitive cell, even mean-field calculations can be computationally very costly.
Several effective Hamiltonian models have been developed to reduce the numerical effort \cite{Gonzalez2017,Vahedi2021,Koshino2018,Song2022}.
Here, we perform a full real-space calculation using the HHK recursive method, which is very efficient to compute the single-particle LDOS of large systems.
The HHK method \cite{Haydock1972, Haydock1975, Haydock1980, Vidarte2022} puts forward a very efficient
Lanczos-like $O({\cal N})$ recursive procedure that transforms an arbitrary sparse Hamiltonian matrix in a tridiagonal one. 
Next, it evaluates the diagonal Green's function in real space by a continued fraction expansion, which is much more numerically amenable than a full diagonalization.

By a suitable choice of the chemical potential, our approach allows us to consider charge neutral as well as systems with a finite doping, namely, $n_{\rm dop}=N_{e}/A_{\rm PUC}$, where $N_{e}$ is the number of electrons in excess to the CNP and $A_{\rm PUC}= 3\sqrt{3}a_{0}^{2}/8\sin^{2}(\theta /2)$ is the area of the TBG moir\'e cell with $N$ atoms. 

We implement the self-consistent mean-field calculation as standard:

($i$) We start with an initial set of occupation numbers $\langle n_{i\sigma}\rangle$, with the constraint $\sum_{i\sigma}\langle n_{i\sigma}\rangle = N + N_e$, where the sum runs over all sites of the moir\'e unit cell, set by the considered doping. The occupations can be chosen randomly or respecting some given symmetry condition. 

($ii$) Using the HHK recursion technique \cite{Vidarte2022,Haydock1972, Haydock1975, Haydock1980}, we compute the LDOS of the electronic system defined by the Hamiltonian, Eq.~\eqref{Eq:Hamiltonian}. 
The spin-resolved LDOS at a given site $i$ can be written as
\begin{align}
\label{eq:LDOS}
\nu_{j}(\epsilon)
% &= -\frac{1}{\pi} {\rm Im}\;G^{r}_{jj}(\epsilon) \\
& = -\frac{1}{\pi} \lim_{\eta \rightarrow 0^{+}}  {\rm Im}\;G_{jj}(\epsilon+i\eta)  .
\end{align}
In practice, the regularization parameter $\eta$ is considered as finite and its magnitude can be attributed the self-energy correction due to disorder, that is ubiquitous in graphene systems.  
%\red
{The HHK method enables us to compute the LDOS with $O(\mathcal{N})$ operations and the DOS with $O(\mathcal{N}^2)$. 
Hence, it is much more efficient than direct diagonalization, which demands $O(\mathcal{N}^3)$ operations.
} 

($iii$) Next, we determine $\langle n_{i\sigma}\rangle$, the average electron occupation number with spin $\sigma$ at the site $i$.
In general, at a given temperature $T$, $\langle n_{i\sigma}\rangle$ reads
\be
    \langle n_{i\sigma}\rangle = \int_{-\infty}^{\infty} {\rm d} \epsilon f_{\mu}(\epsilon) \nu_{i\sigma}(\epsilon) , 
\ee
with $f_{\mu}(\epsilon) = \left\lbrace  \exp{\left[\beta (\epsilon - \mu) \right] } +1 \right\rbrace  ^{-1}$, where $\beta = 1/k_{\rm B}T$, $k_{\rm B}$ is Boltzmann constant, and $\mu$ is chemical potential.
At zero absolute temperature, $\mu$ is equal to the Fermi energy $\ve_F$ and the Fermi distribution becomes a step function.
For simplicity, here we consider $T=0$. 
As standard, the Fermi energy $\ve_F$ (or the chemical potential $\mu$) is determined by the doping. 

($iv$) We examine whether the output occupation numbers $\langle n_{i\sigma}\rangle$ coincide with the input ones, within a tolerance of $10^{-6}$.
If the answer is positive, we end the self-consistent loop.
If not, we return to ($ii$).
The computation of the updated spin densities is then repeated iteratively until all values of $\langle n_{i\sigma}\rangle$ are converged. 

The self-consistent solution provides the spin polarization at the $i$th site,
\begin{align}
\label{Eq: LocalPolarization}
    p_{z,i} &= \dfrac{\langle n_{i\uparrow}\rangle - \langle n_{i\downarrow}\rangle}{2} .
\end{align}
The $p_{z,i}$ is translated in a local magnetization $m_{z,i}=-g\mu_{\rm B}p_{z,i}$, where $\mu_{\rm B}$ is the Bohr magneton and the electron $g$-factor is $g=2$. 
Hence, the total magnetization per moir\'e cell reads
\begin{align}
    M_{\rm PUC}&= \sum^{N}_{i=1} m_{z,i} .
\end{align}

We study two cases: $(i)$ $N_{e}=0$, corresponding to the CNP, and $(ii)$ $N_{e}=3$ corresponding to $3/4$-filling of the conduction miniband \cite{Sharpe2019}.
% The flat bands are four-fold degenerate because of spin and valley symmetries \cite{Cao2016}.
For the CNP case, a previous theoretical study working in reciprocal space has predicted that low-angle TBG systems have an antiferromagnetic ground state \cite{Vahedi2021}.
In this case, we find computationally convenient to start the self-consistent loop with the configuration: $\langle n^{A}_{i\uparrow}\rangle = 1/2 + \Delta n^{A}$, $\langle n^{A}_{i\downarrow}\rangle = 1/2 - \Delta n^{A}$, $\langle n^{B}_{i\uparrow}\rangle = 1/2 + \Delta n^{B}$ and $\langle n^{B}_{i\downarrow}\rangle = 1/2 - \Delta n^{B}$, where $\Delta n^{B} = - \Delta n^{A}$.
For the $3/4$-filling case, we start with a random set of $\langle n_{i\sigma}\rangle$ with the constraint $\sum_{i\sigma} \langle n_{i\sigma}\rangle = N + N_{e}$, where $N_{e} =3$.

%%%%%%%%%%%%%%%%%%%%%%%%%%%%%%%%%%%%%%
\section{Results}
\label{sec:results}
%%%%%%%%%%%%%%%%%%%%%%%%%%%%%%%%%%%%%%

We begin this section by presenting a study of the single-particle LDOS of TBG systems with a focus on the formation of low-energy minibands at small twist angles $\theta$.
Next, we use the obtained LDOS to calculate the magnetization of low-angle TBG systems using the self-consistent procedure outlined in Sec.~\ref{sec:Numericalmethod}.
%We address on two cases: ($i$) TBGs at the CNP and ($ii$) TBGs at $3/4$-filling of the conduction band.

%====================================================================
\subsection{Non-interacting electronic densities}

We compute the local electronic properties of TBGs using the HHK recursion method \cite{Vidarte2022,Haydock1972,Haydock1975,Haydock1980}.
Being a real space approach, the HHK calculations take advantage of the symmetry properties of the odd commensurate TBG structures discussed above.
We have numerically verified that the predicted equivalent sites indeed display the same LDOS.
This is used to reduce the numerical effort by a factor of 6.
For calculational convenience, the regularization factor is set to be  $\eta\approx 5 \dots 25$~meV.
% a parameter range that can be attributed to a small disorder concentration.
We return to this point later on.

\begin{figure}[h!]
\includegraphics[width=0.95\columnwidth]{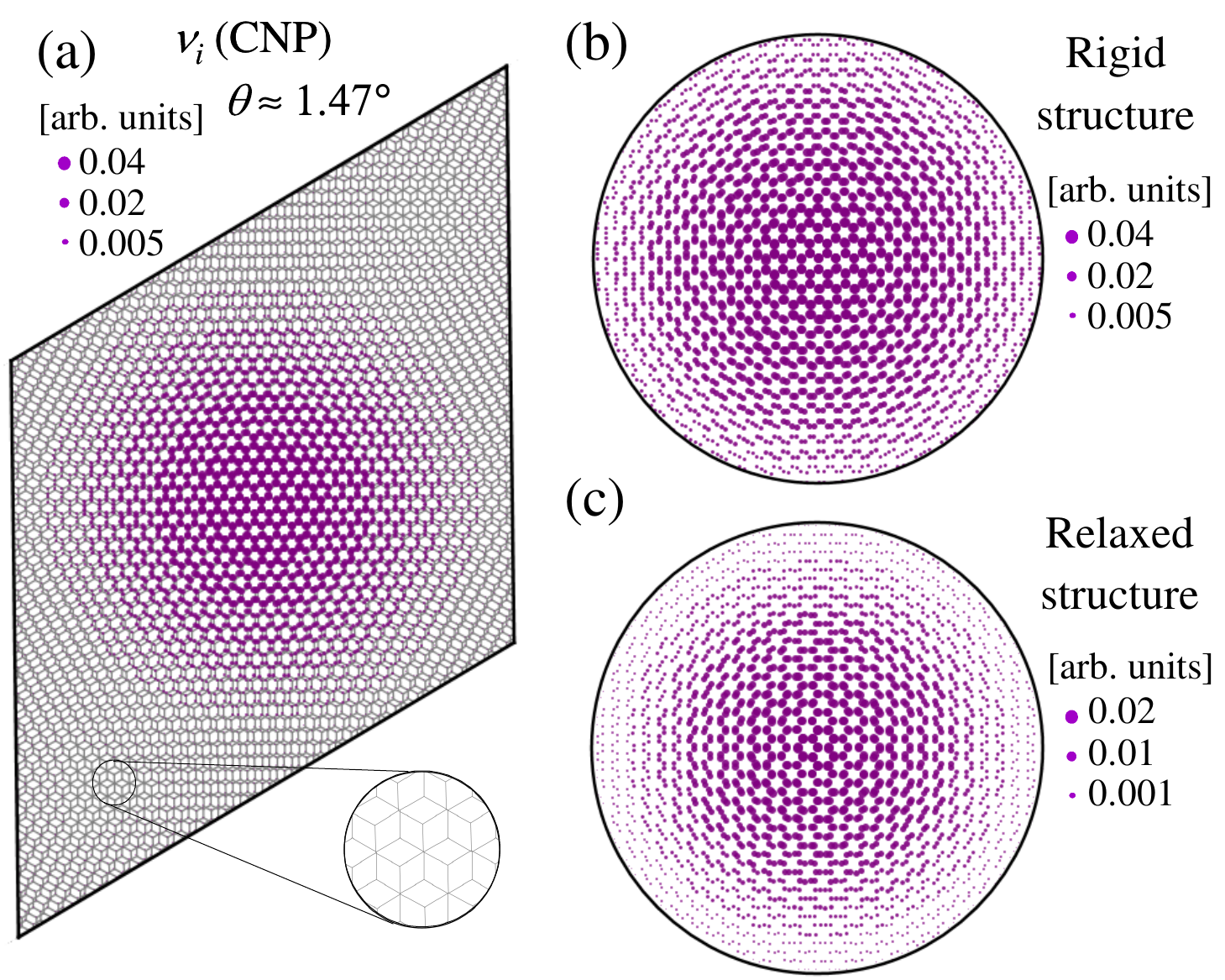}
\caption{Non-interacting electronic densities: 
(a) LDOS at the CNP, $\nu_{i}({\rm CNP})$ (in arbitrary units), for the $(m,n) = (22,23)$ commensurate rigid lattice, that corresponds to $\theta\approx 1.47^{\circ}$ and 6076 atoms within the moir\'e unit cell (black line).
The areas of the purple disks are proportional to $\nu_{i}({\rm CNP})$.
(b) $\nu_{i}({\rm CNP})$ for the same commensurate rigid lattice within the AA-stacking region with a radius of $31.3$~\AA~.
(c) $\nu_{i}({\rm CNP})$ for the relaxed lattice with the same $\theta$ and same circular area of (b).
}
\label{fig:Non-interacting}
\end{figure}

To highlight the sites with the most prominent enhancement of the LDOS in TBGs, we consider the moir\'e region where the AA stacking is placed at the center of the TBG primitive unit cell.
Figure~\ref{fig:Non-interacting}(a) shows the LDOS for a moir\'e unit cell with $(m,n) = (22,23)$ corresponding to a twist angle $\theta\approx 1.47^{\circ}$ containing 6076 carbon atoms.
The AB- (or BA-) stacking region is zoomed in the inset.
Close to the CNP, well-localized states are found in the AA-stacking region, leading to an enhanced LDOS on atoms around the AA stacking, with much smaller LDOS at AB- and BA-stacking regions, in agreement with previous studies \cite{Morell2010, Laissardiere2010}.

%--------------------------------------- F I G U R E -------------------------------------------

%------------------------------------------------------------------------------------------

Figures~\ref{fig:Non-interacting}(b) and \ref{fig:Non-interacting}(c) show the LDOS at the CNP, $\nu_{i}({\rm CNP})$, for rigid and relaxed lattice structures with $\theta\approx 1.47^{\circ}$ within the circular area of radius of $31.3$~\AA,
%that corresponds to the AA-staking region
centered at the AA dimer site (or ${\rm A}^{b}{\rm A}^{t}$ site).
The rigid lattice structure shows a LDOS maximum, $\nu_{\rm AA}({\rm CNP})$, that is twice as large as the one computed for the relaxed lattice.
For both rigid and relaxed lattices, $\nu_{i}({\rm CNP})$ decays radially with respect to the AA dimer site. 
However, the LDOS decays more rapidly for the reconstructed lattices due to in-plane relaxation, which reduces the AA-stacking area.
In contrast, there is no significant contrast in the LDOS at the CNP in the AB-stacking region, $\nu_{\rm AB}({\rm CNP})$, for either rigid or relaxed lattices.

\begin{figure}[h!]
\includegraphics[width=0.95\columnwidth]{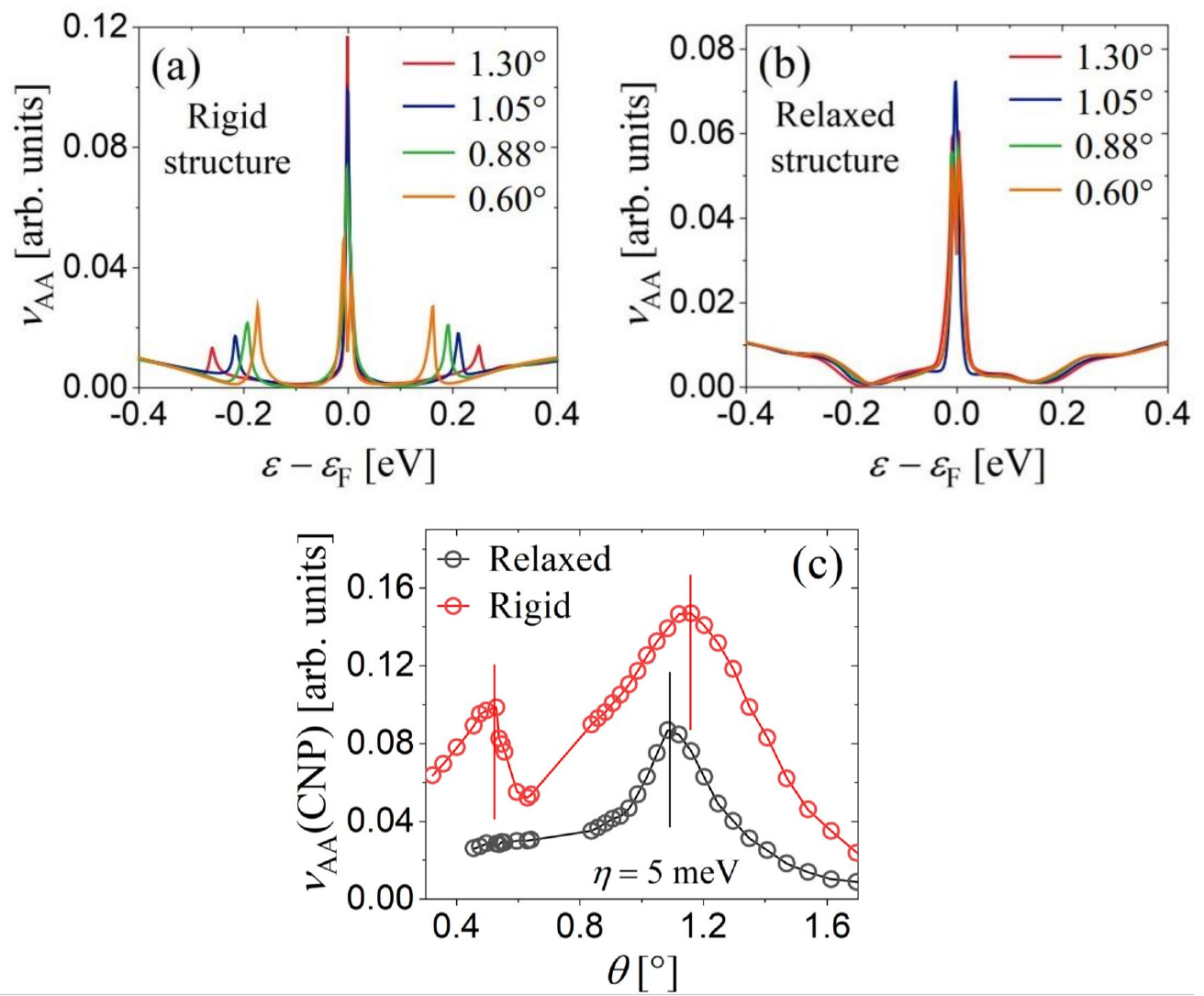}
\caption{
(a) LDOS at the AA dimer sites, $\nu_{\rm AA}$, as a function of the energy (in electron volts) for odd commensurate rigid lattice with $n-m=1$ and $\theta\approx 1.30^{\circ}$, $1.05^{\circ}$, $0.88^{\circ}$ and $0.60^{\circ}$.
(b) $\nu_{\rm AA}$ versus energy for odd commensurate relaxed lattice with the same twist angles $\theta$.
(c) $\nu_{\rm AA}$ at the CNP as a function of $\theta$ (in degrees) corresponding to rigid and relaxed lattice structures with $n-m=1$.
}
\label{fig:LDOS_AAsite}
\end{figure}

Figures \ref{fig:LDOS_AAsite}(a) and (b) show the LDOS at the AA dimer site as a function of the energy $\ve$ around the CNP energy, $\nu_{\rm AA}(\varepsilon)$, for $\theta\approx 1.30^{\circ}$, $1.05^{\circ}$, $0.88^{\circ}$ and $0.60^{\circ}$,  that belong to the family of odd commensurate moir\'e structures with $n-m=1$, for both rigid and relaxed lattices, respectively.
Here we use $\eta = 5$~meV.
The appearance of the central peak can be interpreted in terms of the continuum model \cite{Bistritzer2011}, which associates the enhancement of the LDOS at the vicinity of the magic angle to the formation of a flat band at the CNP.
Figure \ref{fig:LDOS_AAsite}(c) displays the evolution of the AA dimer site LDOS at the CNP as a function of the small twist angles $\theta$ for the $n-m=1$ family of rigid and relaxed lattice structures.

%--------------------------------------- F I G U R E -------------------------------------------
%------------------------------------------------------------------------------------------

Figures \ref{fig:LDOS_AAsite}(a) to (c) show that when the twist angle decreases, the LDOS of the AA-stacking region increases significantly at the CNP.
For the rigid lattice structures, the maximum of the peak appears at the twist angle of $\theta\approx 1.16^{\circ}$ [red vertical line in Fig.~\ref{fig:LDOS_AAsite}(c)] corresponding to $(m,n) = (28,29)$ with 9748 carbon atoms within the moir\'e cell.
For the $n-m=1$ structures considered here, this twist angle is the closest to the first magic angle $\theta^{\rm MA}_{1}\approx 1.1^{\circ}$ predicted by the continuum model \cite{Bistritzer2011, Tarnopolsky2019}.
We find that the flat-band peak width is larger in the relaxed TBG systems than in the rigid counterpart.
The maximum of the LDOS peak appears at $\theta\approx 1.08^{\circ}$ [black vertical line in Fig.~\ref{fig:LDOS_AAsite}(c)], corresponding to $(m,n) = (30,31)$ with 11164 carbon atoms within the moir\'e cell, indicating that lattice relaxation shifts the LDOS peak towards lower twist angles.

Figure~\ref{fig:LDOS_AAsite}(c) shows another peak maximum at the twist angle close to $0.53^{\circ}$ (red vertical line), corresponding to $(m,n) = (62,63)$ with $46876$ carbon atoms within the moir\'e cell for the rigid TBG systems.
This twist angle is the closest to the second magic angle $\theta^{\rm MA}_{2}\approx 0.55^{\circ}$ predicted by the continuum model \cite{Bistritzer2011}.
In Fig.~\ref{fig:LDOS_AAsite}(a), as the twist angle decreases, the satellite peaks around the main one, both in the valence and conduction band, become increasingly intense and move closer to the CNP.
In distinction, for relaxed TBG systems, neither the satellite peaks around the main one nor the second magic angle are observed in the electronic properties of twist angles below $\theta^{\rm MA}_{1}\approx 1.1^{\circ}$ \cite{Nguyen2021,Nguyen2022}. 
In contrast, we find that the van Hove singularities at the AB-stacking region are not depleted, but merely shifted due to lattice relaxation (not shown here), in line with the DOS reported in Ref.~\cite{Nam2017}.

\begin{figure}[h!]
\includegraphics[width=0.95\columnwidth]{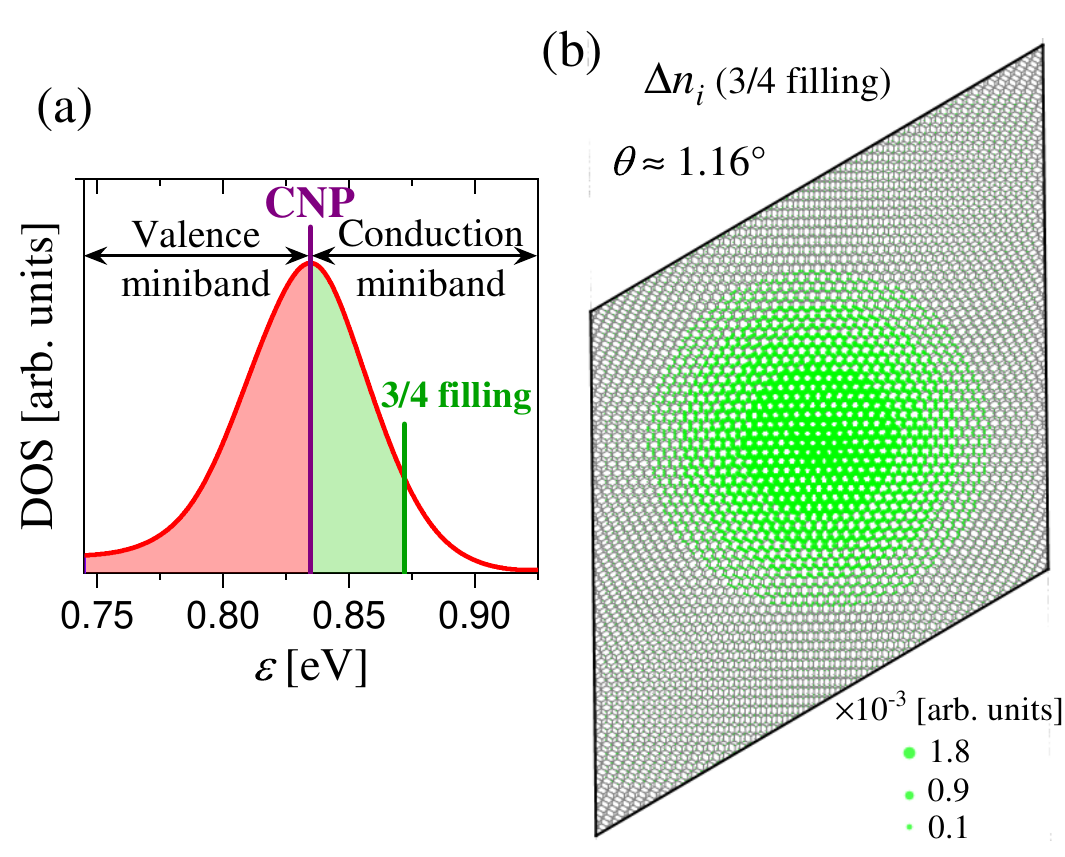}
\caption{Single-particle electron doping of the conduction miniband:
(a) DOS (in arbitrary units) of the flat miniband as a function of the energy (in electron volts).
The CNP corresponds to the DOS peak.
The green area represents 3/4-filling of the conduction miniband.
(b) Carrier distribution within the moir\'e cell (black line) in real space for $\theta\approx 1.16^{\circ}$.
The areas of the green disks are proportional to $\Delta n_{i}$, the occupation in excess to the CNP filling.
}
\label{fig:doping}
\end{figure}

Let us now depart from the CNP and discuss finite doping.
Figure \ref{fig:doping}(a) shows the DOS of the single-particle miniband of a TBG with $\theta\approx 1.16^{\circ}$.
Anticipating the appearance of a gap at the CNP when the interaction is switched on, we define a valence and a conduction miniband separated at the CNP.
The red area represents the half-band filling of a TBG system where the Fermi energy is at the CNP.
This filling corresponds to 4 electrons per moir\'e cell in the valence miniband.
The green area represents the electron doping at the $3/4$ filling of the conduction miniband.
This doping corresponds to 3 electrons per moir\'e cell in excess of the CNP at a Fermi energy $\ve_F = 872$~meV.

%-------------------------------- F I G U R E -------------------------------------------
%------------------------------------------------------------------------------------------

Figure \ref{fig:doping}(b) shows the carrier distribution $\Delta n_{i}$ in excess to the CNP distribution within the moir\'e unit cell for $\theta\approx 1.16^{\circ}$.
The areas of the green disks are proportional to the carrier occupation number at the atomic site $i$, $\Delta n_{i}$.
Owing to the localized LDOS, see Fig. \ref{fig:Non-interacting}, the carrier distribution is also concentrated at the AA-stacking region.

%=================================================================
\subsection{Magnetism at 3/4-filling of the conduction miniband}
\label{sec:3/4filling}
%=================================================================

Let us now examine the ferromagnetic phase in rigid and relaxed TBG systems with a twist angle $\theta\approx 1.16^{\circ}$ at $3/4$-filling of the conduction miniband whose origin has been the subject of discussion in the literature.

We consider an odd commensurate lattice with $(m,n)=(28,29)$ corresponding to 9748 carbon atoms.
Our results are obtained by following the self-consistent procedure described in Sec.~\ref{sec:Numericalmethod} for a finite doping.
Here, we set $U/V^{0}_{pp\pi}=1$.
It is worth to note that we find that a small regularization parameter, such as $\eta=5$~meV, is not essential for obtaining converged results in the interacting case, thereby saving computational cost. 
In our study, we have carefully selected an $\eta$ value that is sufficiently large to ensure accurate integrated DOS and occupation near the Fermi energy, while remaining small enough not to influence the mean-field calculations.
The optimal $\eta$ value may vary between 5 and 25 meV depending on the stacking region considered.

\begin{figure}[h!]
\includegraphics[width=0.95\columnwidth]{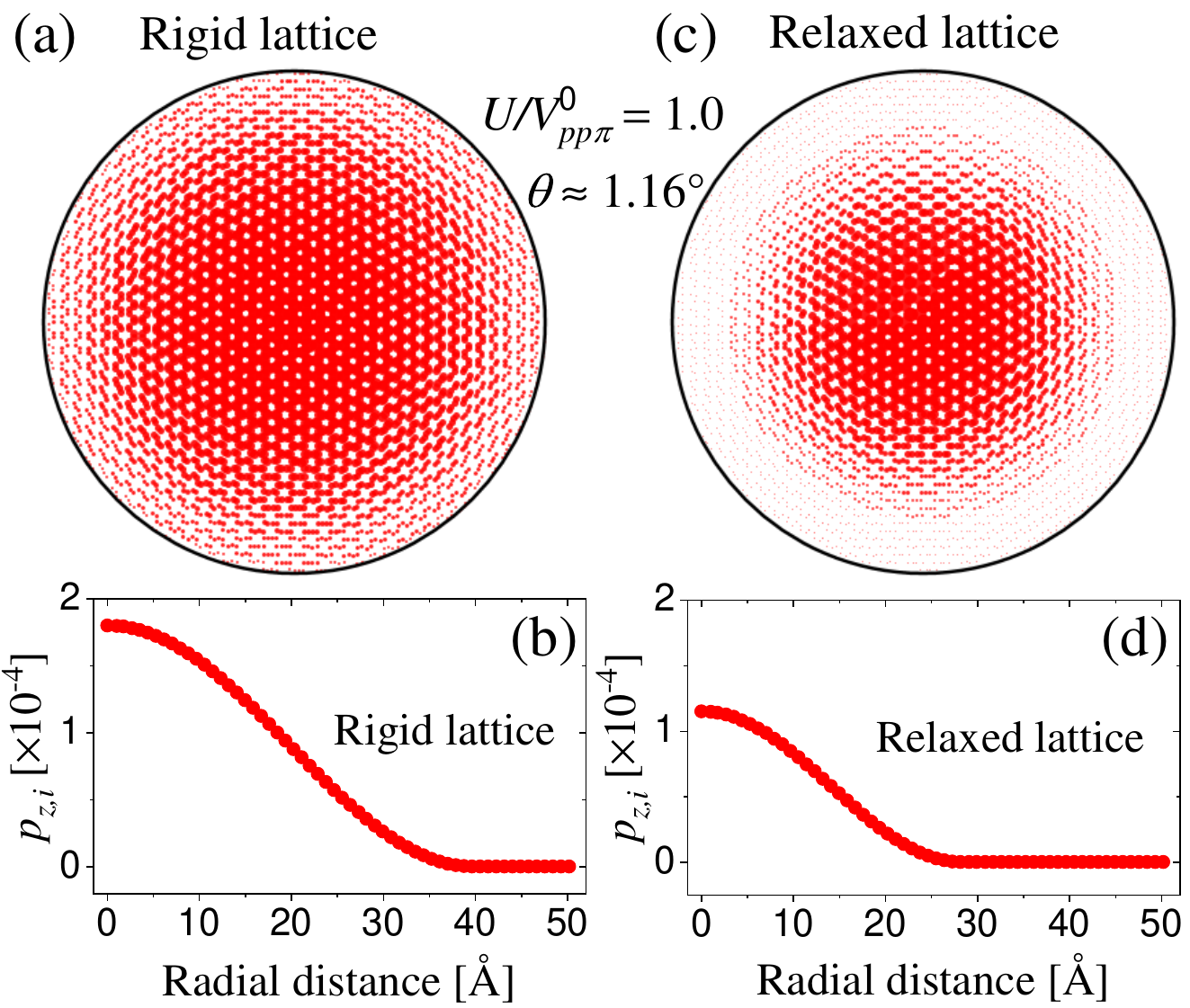}
\caption{Ferromagnetic phase at $3/4$-filling of the conduction miniband:
(a) and (c) Local spin polarization of rigid and relaxed TBG system, respectively, with $\theta\approx 1.16^{\circ}$ and $U/V^{0}_{pp\pi}=1$ within the AA-stacking region with the same circular area.
The areas of the red disks are proportional to the spin polarization $p_{z,i}$.
(b) and (d) Local spin polarization $p_{z,i}$ versus the distance $D$ from the $i$th site to the AA dimer site up to the AB non-dimer site for rigid and relaxed structures, respectively.
}
\label{fig:filling_ConductionMiniband}
\end{figure}

Figures~\ref{fig:filling_ConductionMiniband}(a) and (c) display the local spin polarization, $p_{z,i}$, within a circular area with a radius of $39.9$~\AA~centered at the AA dimer site for rigid and relaxed TBG systems, respectively.
These results show that the ground state is ferromagnetic as characterized by the imbalance between $\langle n_{i\uparrow}\rangle$ and $\langle n_{i\downarrow}\rangle$ in the AA stacking region.
Since the spin polarization is largest at the regions of enhanced LDOS, we attribute the emergence of ferromagnetism to the Stoner instability mechanism. 

%--------------------------------------- F I G U R E -------------------------------------------
%------------------------------------------------------------------------------------------

Figures \ref{fig:filling_ConductionMiniband}(b) and (d) show the local spin polarization, $p_{z,i}$, as a function of the distance $D$ from the AA dimer site for rigid and relaxed TBG systems, respectively.
% Similarly to the CNP case (Appendix~\ref{Appendix}), 
The spin polarization is largest at the AA dimer site and becomes smaller as $D$ increases and eventually vanishes as the $i$th site approaches the AB non-dimer position.
%The magnetization for $U/V^{0}_{pp\pi}=1$ is a small magnitude of the order $10^{-3}$ and is not spatially complete within the moir\'e unit cell for small $U$.

% On the other hand, in distinction to the CNP case (Appendix~\ref{Appendix}), 
Since the ground state is ferromagnetic, the total magnetization is not zero. 
For the rigid lattice structure, we have a maximum local spin polarization of approximately $1.80\times 10^{-4}$ at the AA dimer site, and we find that the total spin polarization per moir\'e cell is $\sim 0.19$ and $M_{\rm PUC} = -0.22$~${\rm eV}\cdot {\rm T}^{-1}$.
However, for the relaxed lattice structure, the maximum local spin polarization is $\sim 1.19\times 10^{-4}$ at the AA dimer site with total spin polarization per moir\'e cell is $\sim 0.13$ and $M_{\rm PUC} = -0.15$~${\rm eV}\cdot {\rm T}^{-1}$.
Furthermore, the local spin polarization decays rapidly within the circular area, see Figs.~\ref{fig:filling_ConductionMiniband}(c) and (d), due to the reduced area of the AA-stacking region.

\begin{figure}[h!]
\includegraphics[width=1.0\columnwidth]{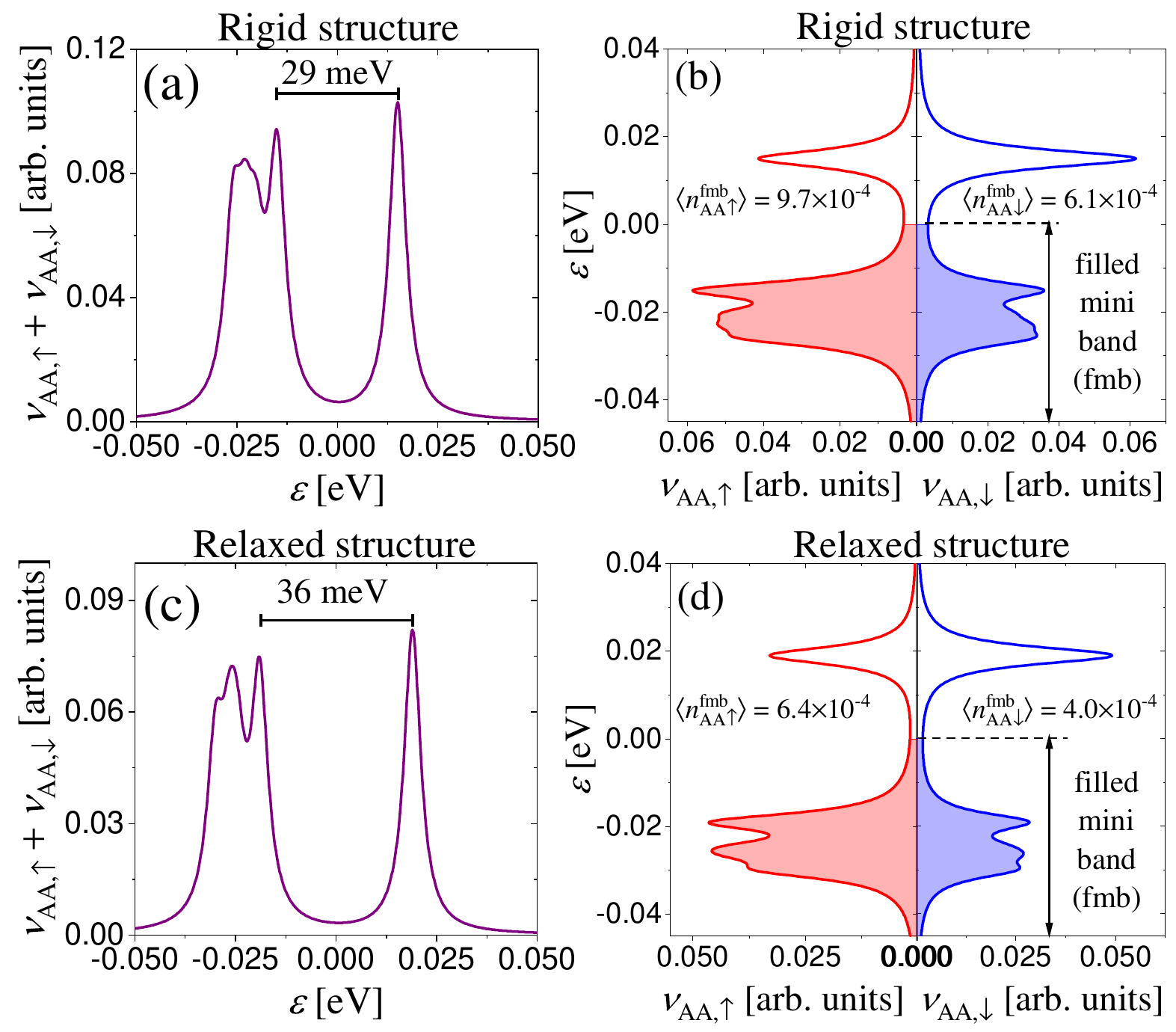}
\caption{
(a) LDOS in the AA dimer site, $\nu_{{\rm AA},\uparrow}+\nu_{{\rm AA},\downarrow}$ (in arbitrary units) as a function of the energy (in eV) for rigid structure.
(b) LDOS for spin up $\nu_{{\rm AA},\downarrow}$ (red line) and spin down $\nu_{{\rm AA},\uparrow}$ (blue line) at the AA dimer site (in arbitrary units) as a function of the energy (in eV) for rigid structure.
(c) $\nu_{{\rm AA},\uparrow}+\nu_{{\rm AA},\downarrow}$ versus energy for relaxed structure.
(d) $\nu_{{\rm AA},\downarrow}$ (red line) and spin down $\nu_{{\rm AA},\uparrow}$ (blue line) at the AA dimer site versus energy for relaxed structure.
From (a) to (d) correspond to $\theta\approx 1.16^{\circ}$ with $U/V^{0}_{pp\pi}=1$.
For convenience, the Fermi energy at $3/4$-filling of the conduction miniband is set to zero.
}
\label{fig:fig_008}
\end{figure}

Figure~\ref{fig:fig_008}(a) and (c) present the LDOS calculated at the AA dimer site, ${\rm LDOS}({\rm AA},\epsilon)$, within an energy window that contains the flat minibands near the Fermi energy (set to $\ve_F=0$ for clarity) for rigid and relaxed structures, respectively. 
We argue that the LDOS(AA,$\ve$) reproduces the behavior of DOS($\ve$) in the vicinity of $\ve_F$, similarly as in the CNP case studied in Ref.~\cite{Vahedi2021} (see, App.~\ref{Appendix}). 
% The LDOS at the AA dimer site shows three LDOS peaks.
% We note that the two LDOS peaks of the energetically lowest miniband contain a larger number of charge % carriers than the LDOS peak of the conduction miniband, due to the enhanced LDOS in the AA stacking 
% region being doped by electrons.
The separation of the two LDOS peaks around the Fermi energy is approximately $\Delta \approx 29$~meV for a rigid lattice structure and $\Delta \approx 36$~meV for a
relaxed one.
We speculate that this modest enhancement is due to the fact that, despite the slightly smaller DOS observed in the relaxed structure, the reduced size of AA-stacking region  amplifies Coulomb repulsion effects. 
We find that the gap sizes are spatially constant within the AA-stacking region.
The existence of two LDOS peaks in the valence miniband suggests the existence of two minibands that overlap in the CNP, as indicated in Ref.~\cite{Cao2018a}.

%--------------------------------------- F I G U R E -------------------------------------------
%------------------------------------------------------------------------------------------

Figure \ref{fig:fig_008}(b) depicts the spin up and spin down LDOS at the AA dimer site, denoted as $\nu_{{\rm AA},\uparrow}(\epsilon)$ and $\nu_{{\rm AA},\downarrow}(\epsilon)$, respectively, for rigid lattice structure.
A clear spin-imbalance is evident between the spin-up and spin-down occupation numbers in the filled miniband (fmb).
Quantitatively, the occupation numbers are $\langle n^{\rm fmb}_{{\rm AA},\uparrow}\rangle = 9.7\times 10^{-4}$ and $\langle n^{\rm fmb}_{{\rm AA},\downarrow}\rangle = 6.1\times 10^{-4}$.
This imbalance translates to a spin polarization $ p_{z,{\rm AA}} = 1.8\times 10^{-4}$ at the AA dimer site, corresponding to a local magnetic moment  $m_{z,{\rm AA}}= -0.21$~${\rm meV}\cdot {\rm T}^{-1}$.
In Fig.~\ref{fig:fig_008}(d), which corresponds to the relaxed lattice structure, the occupation numbers are $\langle n^{\rm fmb}_{{\rm AA},\uparrow}\rangle = 6.4\times 10^{-4}$ and $\langle n^{\rm fmb}_{{\rm AA},\downarrow}\rangle = 4.0\times 10^{-4}$, with a spin polarization $ p_{z,{\rm AA}} = 1.2\times 10^{-4}$ at the AA dimer site.

%\red{
In summary, the Stoner mechanism is robust against the suppression of the flat-band LDOS peak due to lattice relaxation. 
The ferromagnetic phase is preserved, but the magnitude of the local magnetic moments becomes smaller than the ones computed for rigid structures.
%}

%%%%%%%%%%%%%%%%%%%%%%%%%%%%%%%%%%%%%%
\section{Conclusions}
\label{sec:conclusion}
%%%%%%%%%%%%%%%%%%%%%%%%%%%%%%%%%%%%%%

In this work, we investigate the emergence of magnetism in low-angle TBG systems using a numerical real-space approach.
We have considered a tight-binding Hamiltonian with a mean-field Hubbard term and obtained the ground state of TBG systems by a self-consistent iteration procedure.
Notably, owing to the HHK recursive technique \cite{Vidarte2022,Haydock1972, Haydock1975, Haydock1980}, our approach allows calculations to be efficiently performed for very large moir\'e cells.
%\ERIC{I don't know if it is convenient to put 1st paragraph on page 7, right column in the conclusions. I would put it before.}
%\CAIO{Kevin and I did not quite understand what you propose. The way it is now, we begin the conclusions with a summary of what is done. For a long paper, I believe this is OK.}
%\ERIC{Ok. Not a big issue. We can leave it here.}
To validate our method, as detailed in App.~\ref{Appendix}, we compare our results for low-angle TBGs at the CNP with those previously reported in the literature \cite{Vahedi2021}.
Specifically, we have found the emergence of an antiferromagnetic phase for low-angle TBGs with a maximum local spin polarization $p_{z,i}$ near the magic angle.
For all computed twist angles $\theta$, the agreement of $p_{z,i}$ with Ref.~\cite{Vahedi2021} is excellent, demonstrating the accuracy of our computational procedure.

The main finding of this study is the emergence of a ferromagnetic phase in low-angle TBGs at $3/4$-filling of the conduction miniband.
Motivated by recent experiments \cite{Sharpe2019,Sharpe2021} we have investigated the DOS, the local and total magnetization of TBGs for the rotational angle $\theta\approx 1.16^{\circ}$.
Our calculations for the case of $3/4$-filling of the conduction miniband (3 electrons in excess to the CNP per moir\'e cell) show that the system ground state is a ferromagnetic insulator, in agreement with the experimental findings \cite{Sharpe2019}, with a gap of $\Delta\approx 36$~meV.
This gap is larger than the one experimentally reported in MATBGs at the charge neutrality point, but one should keep in mind that these are different systems with distinct underlying physics, which may account for the observed discrepancies.
We attribute this behavior to the large LDOS at the AA-stacking region, as predicted by the continuum \cite{Bistritzer2011,Tarnopolsky2019} and tight-binding model \cite{Laissardiere2010,Laissardiere2012}, that causes a strong enhancement of the electron-electron interaction, a key element for the Stoner mechanism.
%\red{
Our method neither challenges the substrate-induced gap proposed in Ref.~\cite{Zhang2019, Bultinck2020}, nor addresses the chiral edge observed in Ref.~\cite{Sharpe2019}.
Instead, we provide an alternative scenario for the appearance of the ferromagnetic phase, regardless of the substrate alignment, which can serve as a motivation for additional experimental studies.
%}

We expect our methodology to be useful for studying the interaction effects of other moir\'e 2D systems with large primitive unit cells.
For instance, with modest adaptations, our approach can be used in the analysis of anisotropic ferromagnetism dominated by the orbital magnetic moment in 3/4-filling at the conduction miniband of MATBG systems \cite{Sharpe2021}.
%\red{\sout{
%By combining our calculations with with a molecular dynamics procedure, we plan to analyze lattice relaxation effects as a function of twist angle. 
%}}
In addition, the presence of an external magnetic field that typically requires the use of large supercells in standard approaches is trivially accounted for in our method.
%\red{
Finally, an important development of our method for low-angle TBGs is the inclusion of Coulomb interactions between the localized electronic states, which is the next goal we plan to pursue.
%}

\appendix
%%%%%%%%%%%%%%%%%%%%%%%%%%%%%%%%%%%%%%
\section{Magnetism at the CNP}
\label{Appendix}
%%%%%%%%%%%%%%%%%%%%%%%%%%%%%%%%%%%%%%

Here, we compare the results obtained using the self-consistent scheme described in Sec.~\ref{sec:Numericalmethod} with those presented in Ref.~\cite{Vahedi2021} (that considered rigid structures only).
The excellent agreement serves to validate our method. 
% our approach Using the mean-field Hubbard Hamiltonian, we implement the self-consistent scheme described in Sec.~\ref{sec:Numericalmethod} to study the emergence of antiferromagnetic states in low twist angle TBGs at the CNP. 
% and compare our results with Ref. \cite{Vahedi2021}.

\begin{figure}[h!]
\includegraphics[width=0.95\columnwidth]{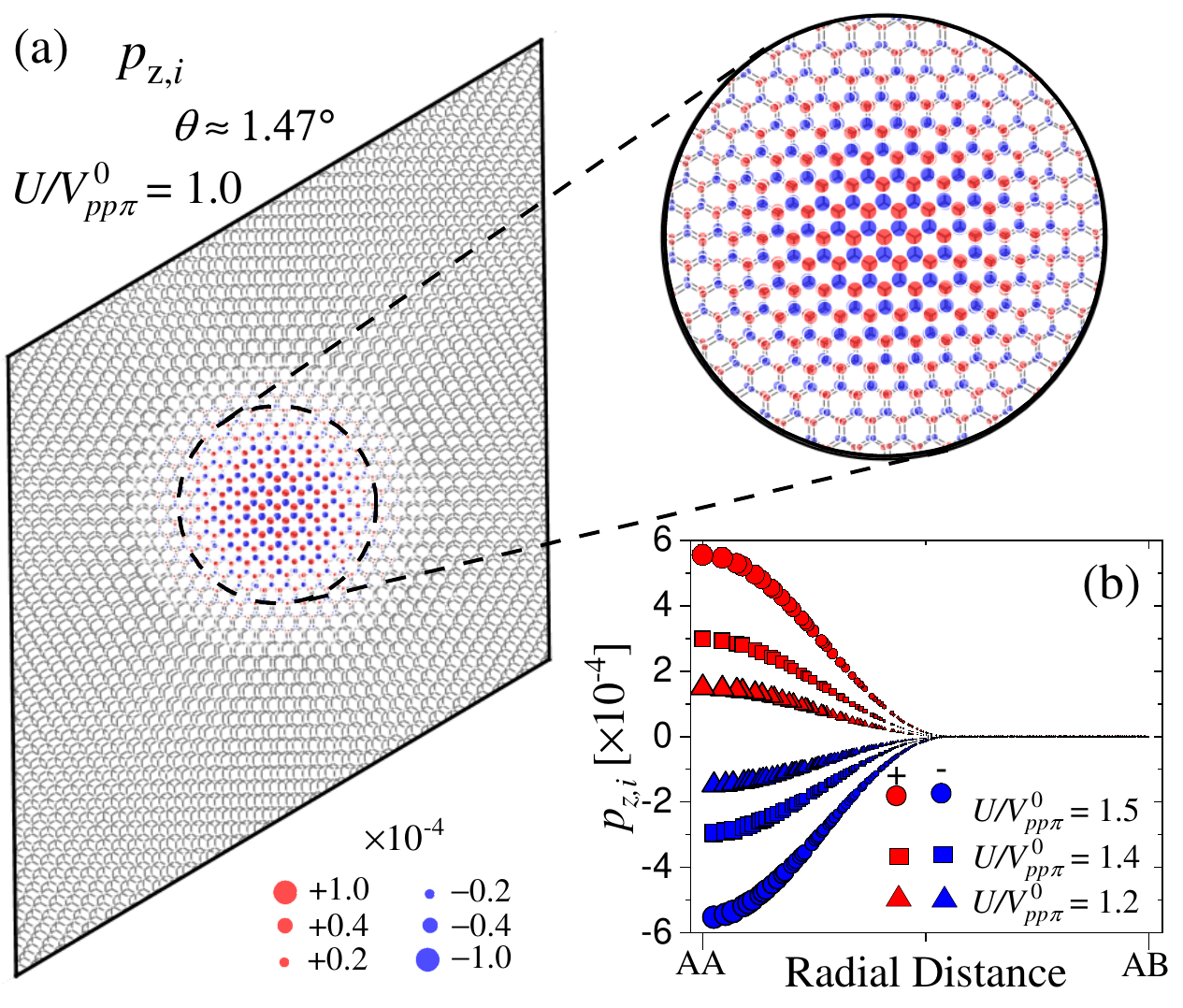}
\caption{Antiferromagnetic phase at the CNP:
(a) Local spin polarization $p_{z,i}$ within the moir\'e cell (black line) for a TBG with $\theta \approx 1.47^{\circ}$ and $U/V^{0}_{pp\pi}=1$.
The disk areas are proportional to $\vert p_{z,i}\vert$, where the red and blue disks correspond to positive and negative spin polarizations, respectively.
(b) Local spin polarization $p_{z,i}$ as a function of the radial distance with respect to the AA dimer site (or AB non-dimer site) for $U/V^{0}_{pp\pi}=1.2$, $1.4$ and $1.5$ with $\theta \approx 1.47^{\circ}$.
}
\label{fig:Antiferro_01}
\end{figure}

Figure \ref{fig:Antiferro_01}(a) displays the local spin polarization, $p_{z,i}$, within the moir\'e unit cell for a $(m,n)=(22,23)$ lattice corresponding to a rotation angle $\theta \approx 1.47^{\circ}$ with $U/V^{0}_{pp\pi}=1$.
The regions with the largest local magnetic moments correspond to those of enhanced LDOS, see Fig.~\ref{fig:Non-interacting}, strongly suggesting that the emergence of magnetism can be attributed to the Stoner mechanism \cite{Auerbach1998}. 
In the AA stacking region, the imbalance between $\langle n_{i\uparrow}\rangle$ and $\langle n_{i\downarrow}\rangle$ leads to the emergence of an antiferromagnetic ground state at the CNP, as can be seen in the zoom. 
%\sout{We note in passing that the converged solution is the same if the we start the self-consistent loop with a distribution of random or antiferromagnetic occupations $\langle n_{i\sigma}\rangle$.}

%--------------------------------------- F I G U R E  4 -------------------------------------------
%------------------------------------------------------------------------------------------

Figure~\ref{fig:Antiferro_01}(b) shows the local spin polarization for a TBG system with $\theta \approx 1.47^{\circ}$ for different values of the on-site Coulomb interaction, $U/V^{0}_{pp\pi}=1.2$, $1.4$ and $1.5$, as a function of the radial distance to the AA dimer site. 
% (or to the AB non-dimer site).
Here, the distance between the AA dimer site and the AB nondimer one is $D = 55.3 $~\AA.
In line with Fig.~\ref{fig:Antiferro_01}(a), one observes that the spin polarization shows a maximum at the AA-stacking region and gradually decreases and eventually vanishes for the sites that are closer to the AB stacking.
The magnitude of local spin polarization becomes increasingly larger with increasing on-site Coulomb interaction $U$, and the overall dependence on the radial distance to the AA dimer site is similar for all $U$ values considered.
Note that since the ground state is antiferromagnetic, the total spin polarization per moir\'e cell is zero for all values of $U$.
Figure~\ref{fig:Antiferro_01} shows excellent agreement with the reciprocal space calculation reported in Ref.~\cite{Vahedi2021}.

\begin{figure}[h!]
\centering
\includegraphics[width=0.70\columnwidth]{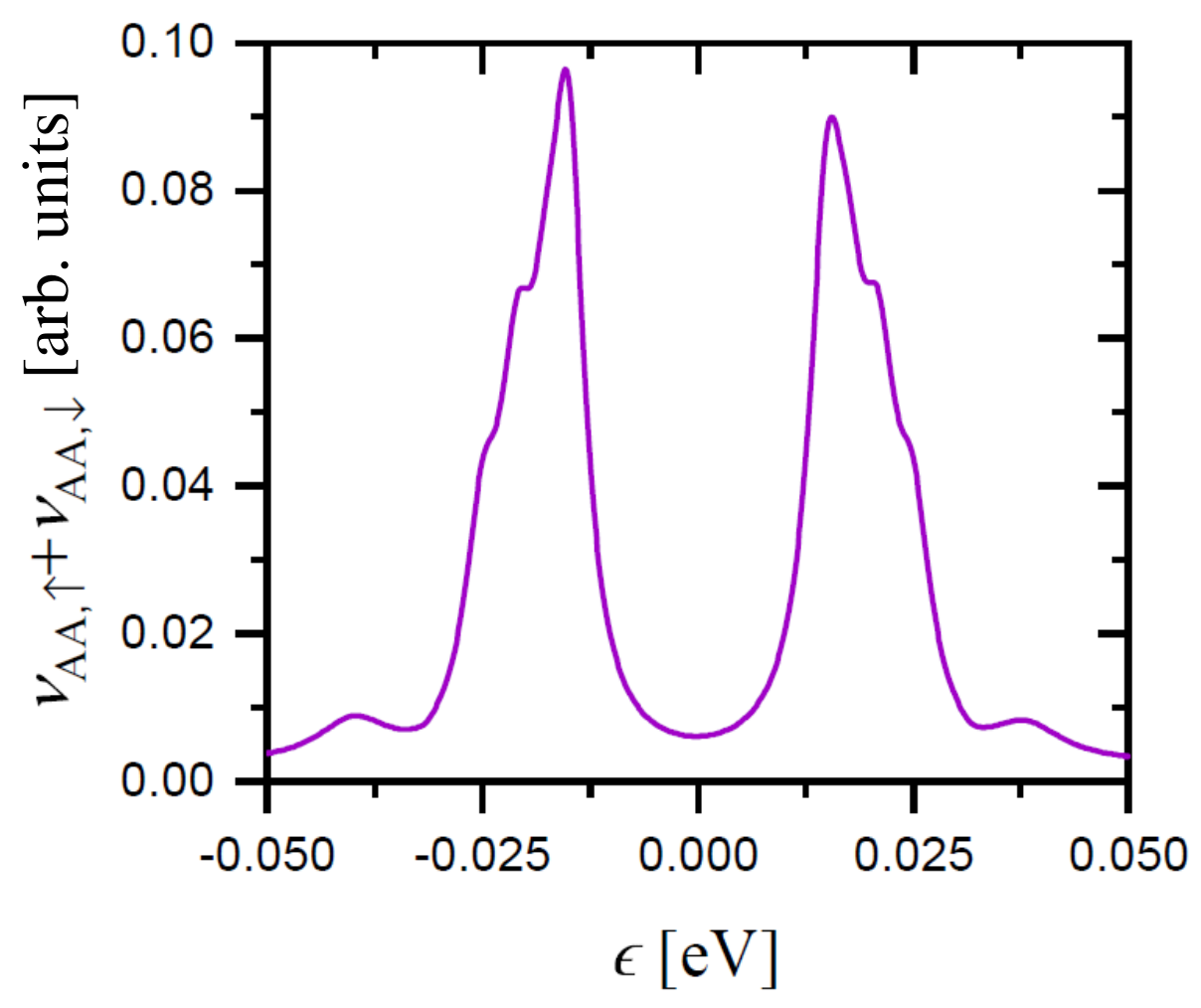}
\caption{
LDOS at the AA dimer site, $\nu_{{\rm AA},\uparrow}+\nu_{{\rm AA},\downarrow}$ (in arbitrary units) as a function of energy $\ve$ (in eV) for $\theta\approx 1.47^{\circ}$ with $U/V^{0}_{pp\pi}=1$.
For convenience, the Fermi energy at the CNP is set to zero.
}
\label{fig:Fig_08}
\end{figure}

Figure~\ref{fig:Fig_08} displays the LDOS of the flat miniband calculated for AA dimer site, namely,  ${\rm LDOS}({\rm AA},\epsilon)=\nu_{{\rm AA},\uparrow} (\epsilon)+\nu_{{\rm AA},\downarrow}(\epsilon)$ for a twist angle of $\theta\approx 1.47^{\circ}$ with  $U/V^{0}_{pp\pi}=1$.
Due to the electron-electron interaction, the LDOS of the flat miniband in the AA dimer site is split into two peaks almost symmetric around the Fermi energy, here set to $\epsilon_{F}=0$, and a small gap is opened.
The separation between the two LDOS peaks is approximately $0.031$~eV.
Note that since the charge density is not homogeneous, one expects ${\rm LDOS}({\rm AA},\epsilon)$ to differ from the  DOS($\ve$).
However, since the two low-energy LDOS peaks are absent in the atomic sites at the AB-stacking region, the ${\rm LDOS}({\rm AA},\epsilon)$ captures the main features of ${\rm DOS}(\epsilon)$ at low energies.

%--------------------------------------- F I G U R E -------------------------------------------
%------------------------------------------------------------------------------------------

\begin{figure}[h!]
\includegraphics[width=0.95\columnwidth]{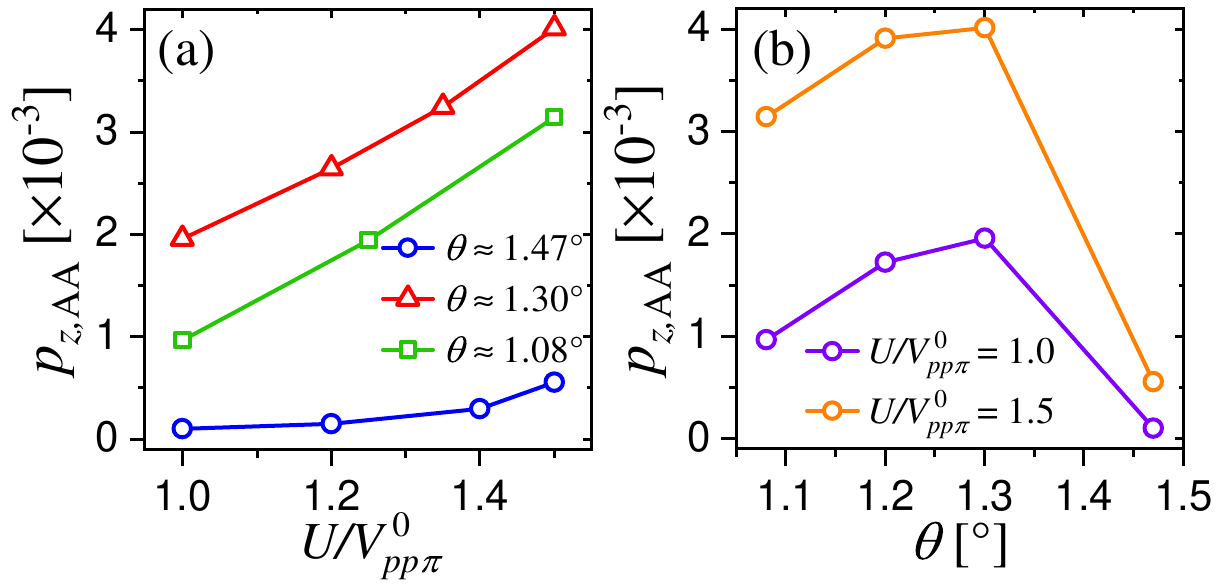}
\caption{
(a) Local spin polarization at the AA dimer site versus $U/V^{0}_{pp\pi}$ for $\theta \approx 1.47^{\circ}$, $1.30^{\circ}$ and $1.08^{\circ}$.
(b) Local spin polarization at the AA dimer site as a function of the twist angle $\theta$ (in degrees) for $U/V^{0}_{pp\pi} = 1.0$ and $1.5$.
}
\label{fig:Antiferro_02}
\end{figure}

Figure \ref{fig:Antiferro_02}(a) shows the maximum spin polarization that corresponds to the AA dimer site, $p_{z,{\rm AA}}$, as a function of the on-site Coulomb repulsion $U/V^{0}_{pp\pi}$ for different twist angles $\theta \approx 1.47^{\circ}$, $1.30^{\circ}$ and $1.08^{\circ}$.
We observe that the local spin polarization at the AA dimer site increases monotonically with increasing $U$.
Furthermore, $p_{z,{\rm AA}}$ versus $U/V^{0}_{pp\pi}$ shows a similar (almost linear) slope for $\theta\approx 1.08^{\circ}$ and $\theta\approx 1.30^{\circ}$, while for $\theta\approx 1.47^{\circ}$ both $p_{z,{\rm AA}}$ and its slope with respect to $U/V^{0}_{pp\pi}$ are much smaller.
This suggests that $\theta\approx 1.47^{\circ}$ is close to a critical angle where the magnetic phase ceases to exist.
A study of the critical values of $U_{c}/V^{0}_{pp\pi}$ for the zero temperature normal antiferromagnetic transition of TBG systems at the CNP can be found in Ref.~\cite{Vahedi2021}, which estimates the critical values of $U_{c}/V^{0}_{pp\pi}$.

%--------------------------------------- F I G U R E -------------------------------------------
%------------------------------------------------------------------------------------------

Figure \ref{fig:Antiferro_02}(b) shows the maximum spin polarization, $p_{z,{\rm AA}}$, as a function of the magic angle $\theta$ for $U/V^{0}_{pp\pi} = 1.0$ and $1.5$.
We note that $p_{z,{\rm AA}}$ shows a maximum close to $\theta = 1.30^{\circ}$ for different values of $U/V^{0}_{pp\pi}$, consistent with Fig.~\ref{fig:Antiferro_02}(a).
This value of $\theta$ is slightly shifted with respect to the magic angle, $\theta_{\rm MA}\approx 1.1^{\circ}$, which can be attributed to small differences in the corresponding single-particle model Hamiltonian parameterizations.

%-----------------------------------------------------------------------
\acknowledgments
We thank J. Vahedi for the discussions and for providing the data of his computations of TBG systems.
This work was partially supported by the Brazilian Institute of Science and Technology (INCT) in Carbon Nanomaterials and the Brazilian agencies CAPES, CNPq, FAPEMIG, and FAPERJ.
ESM acknowledges financial support from ANID Chile Fondecyt 1221301.

%-----------------------------------------------------------------------
\bibliography{methods,hhk,TBG,magnetism-graphene}
%-----------------------------------------------------------------------
\end{document}